# High-frequency performance of graphene field effect transistors with saturating IV-characteristics

Inanc Meric[1], Cory R. Dean[1,2], Shu-Jen Han[3], Lei Wang[2], Keith A. Jenkins[3], James Hone[2], and K. L. Shepard[1]

[1]Department of Electrical Engineering, [2]Department of Mechanical Engineering, Columbia University, New York, NY, 10027
[3]IBM T.J. Watson Research Center, Yorktown Heights, NY 10598
Tel: (646) 205-0438, Fax: (212) 932-9421, Email: shepard@ee.columbia.edu

**Abstract**

High-frequency performance of graphene field-effect transistors (GFETs) with boron-nitride gate dielectrics is investigated. Devices show saturating IV characteristics and $f_{max}$ values as high as 34 GHz at 600-nm channel length. Bias dependence of $f_T$ and $f_{max}$ and the effect of the ambipolar channel on transconductance and output resistance are also examined.

## Introduction

Interest remains high in the potential use of graphene as a field-effect transistor (FET) channel replacement material [1, 2]. The focus is primarily on analog and RF applications of graphene FETs (GFETs) because of the limited on-current-to-off-current ratios achievable with this zero-bandgap material. Within the last few years, the RF performance of GFETs, as determined by the device current-gain cut-off frequency ($f_T$), has gone from 15 GHz [3] for 500-nm-length devices in the first measurements to 155 GHz at 40-nm channel lengths in the most recent reports [4].

RF measurements have generally been reported for top-gated device structures whose current-voltage characteristics do not show strong current saturation due to relatively poor gate-oxide interfaces or weak gate coupling. As a result, device output conductance is high, power gain is limited, and the maximum oscillation frequency ($f_{max}$) is typically only one-tenth of $f_T$. In this work, by exploiting high-quality boron-nitride dielectrics, we instead find $f_{max}/f_T$ ratios as high as 0.86 and $f_{max}$ values as high as 34 GHz for a 600-nm-length device, the highest value reported so far for GFETs. We further investigate the bias dependence of both $f_T$ and $f_{max}$ and compare our results with small-signal models of our device structures.

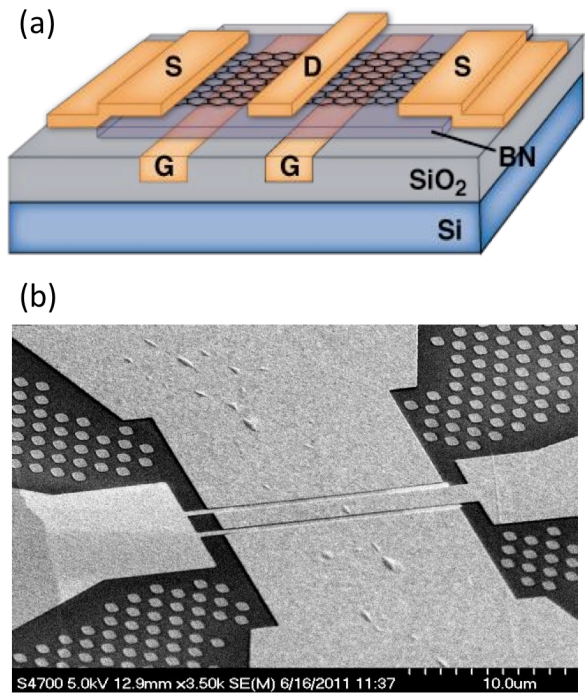

**Figure 1. GFET device structure.** (a) Schematic illustration of the back-gated GFET device. (b) SEM micrograph of a completed structure.

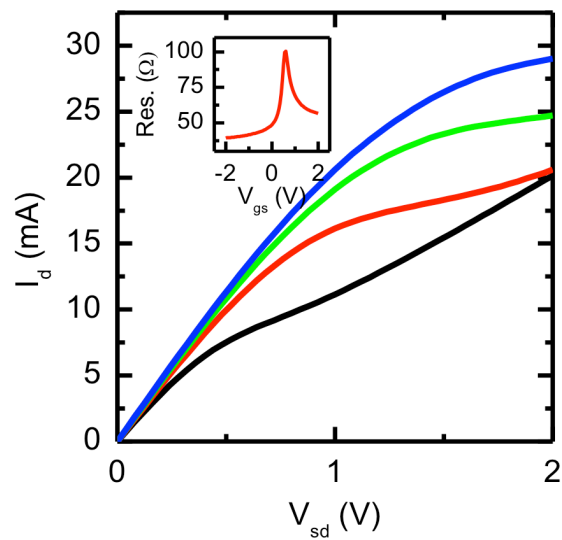

**Figure 2. IV characteristics.** $I_D$ as a function of $V_{sd}$ for $V_{sg}$ from 0V to -1.5V in 0.5V steps. <u>Inset:</u> Resistance in linear transport region as a function of $V_{gs}$.

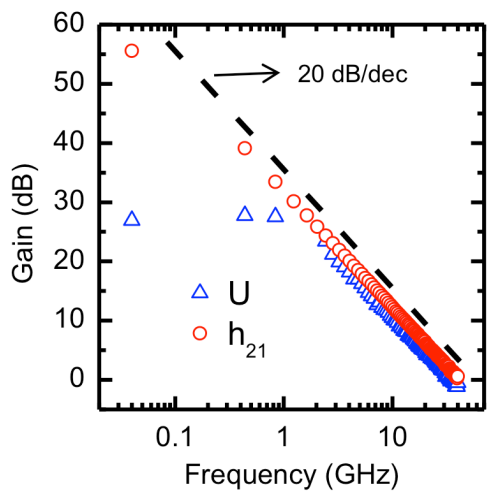

**Figure 3. High-frequency device characteristics.** $h_{21}$ and unilateral power gain (U) as a function of frequency after deembedding, yielding $f_T$ = 44 GHz and $f_{max}$ = 34 GHz.

## Device Fabrication

Hexagonal boron nitride (h-BN) has been previously found to be an outstanding gate dielectric for GFETs, yielding interfaces nearly free of trapped charge and maintaining high mobility and carrier velocities in the graphene channel [5, 6]. The GFETs characterized here are created with a back gate as shown in Fig. 1a. A split-gate layout is employed, where tungsten metal gates are initially patterned into a 1-μm $SiO_2$ layer using a Damascene-like process, followed by a chemical-mechanical polishing (CMP) step to ensure a flat surface and expose the gate metal surface. h-BN (10-nm thick) is mechanically transferred to form the gate dielectric, followed by the mechanical transfer of the graphene channel (single layer). GFET fabrication ends with e-beam patterning of source and drain contacts with approximately 50-nm gate-to-source and gate-to-drain spacings as shown in Fig. 1a. An SEM micrograph of a completed device is shown in Fig.1b.

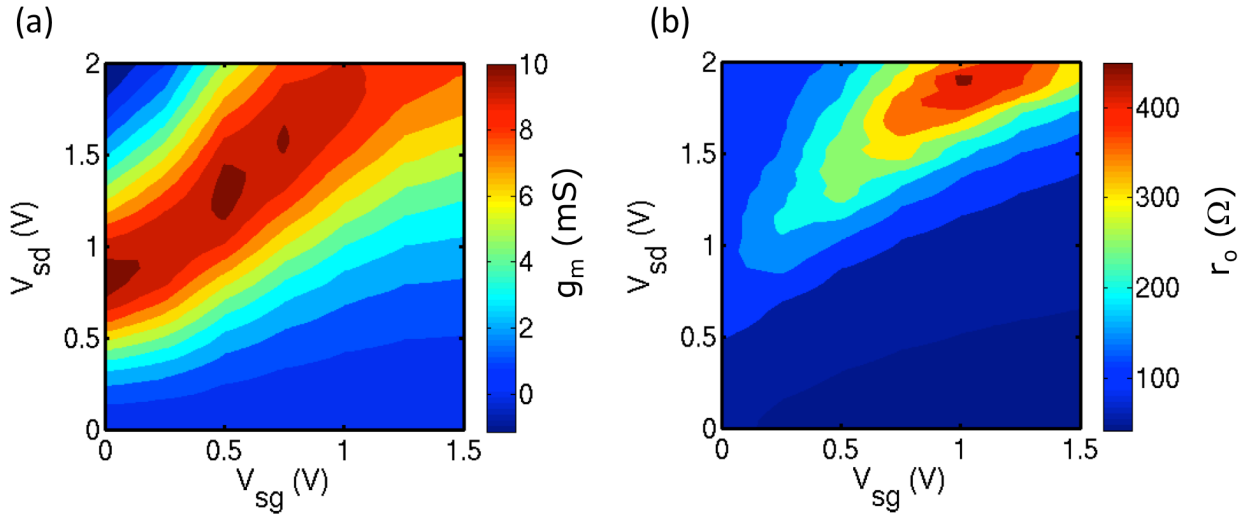

**Figure 4. Bias-dependence of small-signal parameters. (a)** Transconductance and **(b)** output resistance.

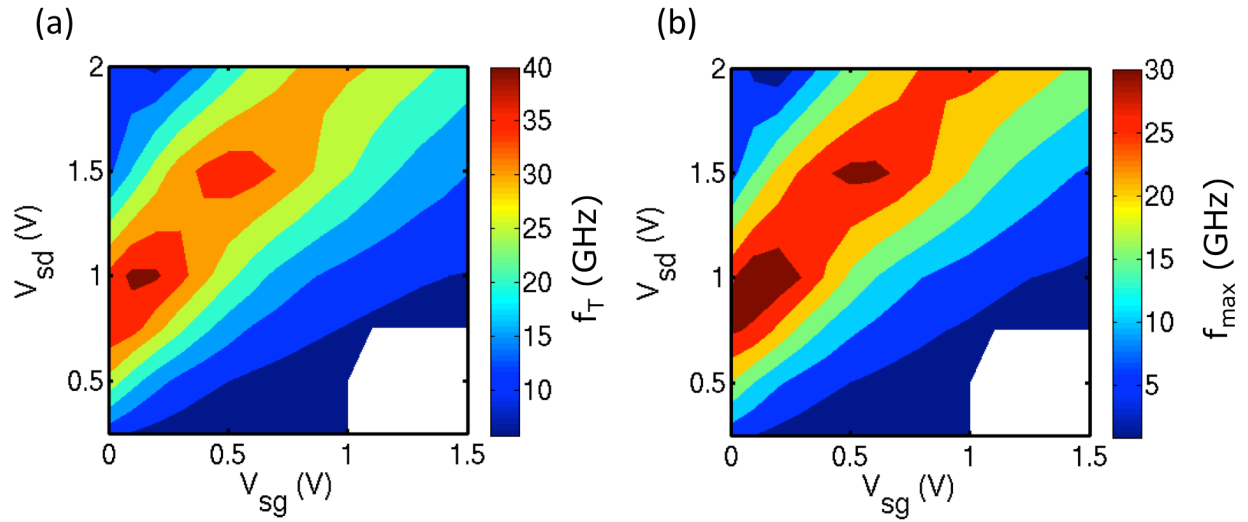

**Figure 5. Bias dependence of high-frequency figures-of-merit.** (a) $f_T$ and (b) $f_{max}$.

## DC Measurements

Fig. 2 shows the DC current-voltage (IV) characteristic of a representative GFET device with an effective width of approximately 38 μm and channel length of 0.6 μm. The inset of Fig. 2 shows the accompanying source-drain resistance in the triode region at $V_{sd}$ = 10 mV, from which the contact resistance and low-field mobility can be extracted. The total contact resistance (including both source and drain) is approximately 25 Ω, or 950 Ω-μm when normalized to contact width. (Contact resistance is inversely proportional to contact width.) The low-field mobility is 3,300 cm$^2$/V sec. The charge neutrality point ($V_o$), the gate-to-source voltage at which the maximum low-field triode resistance is achieved, is 0.6 V. IV characteristics are plotted for gate voltages ($V_{sg}$) from 0 to -1.5V, demonstrating both saturating current characteristics for the unipolar hole channel and the "kink" associated with the transition to the ambipolar hole-electron channel. Fig. 4 gives the transconductance ($g_m$) and output resistance ($r_o$) as a function of bias; both are strongly influenced by the kink behavior in the IV characteristic. The peak $g_m$ of 10.5 mS and peak $r_o$ of 460 Ω occur at different bias points. The devices show no hysteresis and unchanged characteristics after repeated measurements.

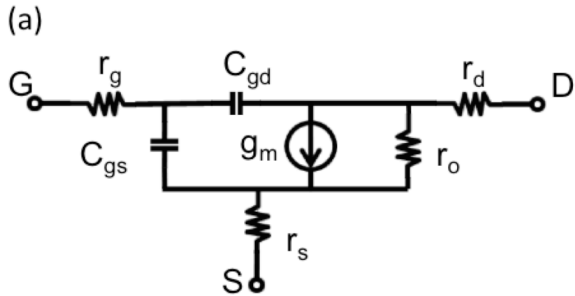

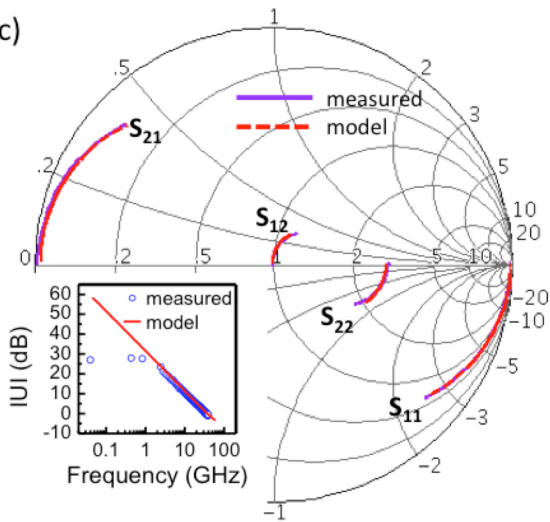

Figure 6. **Small-signal modeling.** (a) Small-signal equivalent circuit for a representative 0.6-μm transistors; (b) parameter values used in the model; and (c) measured S-parameters along with the results of the model. Inset: Unilateral gain of both measured and modeled device.

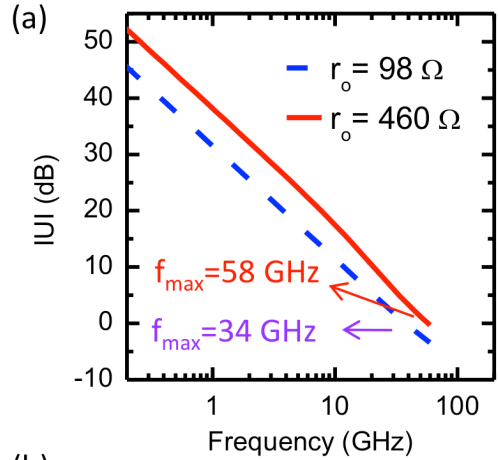

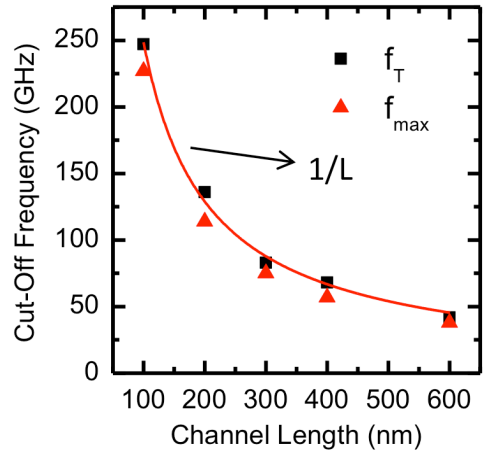

Figure 7. **Improvements possible with device optimization.** (a) U calculated from the model with parameter values of Fig. 6b (dashed blue) and with output resistance set to highest measured value (red). (b) Predicted high-frequency performance at different channel lengths.

## High-Frequency Measurements

Device S-parameters are measured to 40 GHz. Standard "open-short" de-embedding methods are employed. In Fig. 3, current-gain ($h_{21}$) and unilateral power gain ($U$) are plotted at the bias point of peak $g_m$, yielding $f_T$ and $f_{max}$ of 44 GHz and 34 GHz, respectively. (Without de-embedding $f_T$ and $f_{max}$ are 24 GHz and 17 GHz, respectively.) Fig. 5 shows $f_T$ and $f_{max}$ as a function of bias; the peak high-frequency response closely matches the peak transconductance of the device.

The small-signal model of Fig. 6a is used to model the high-frequency behavior of the GFETs. The measured S-

parameters are shown in Fig. 6c, showing good agreement with the results of the small signal model with the parameters given in Fig. 6b. These small-signal values are in good agreement with values derived from the IV characteristics. The $g_m$ here is the intrinsic value, exclusive of contact resistance, in good agreement with our previous results of approximately 0.5 mS/µm [6].

The model derived from this representative 0.6-µm device can be used to further understand device optimization and scaling. Fig. 7a shows how the $f_{max}$ performance could be improved to 58 GHz for this same channel length if the $V_o$ of the device could be adjusted (through a secondary gate or channel doping) to align peak $g_m$ and $r_o$. The model is also used to estimate the performance at shorter channel lengths by scaling gate capacitance while keeping other small-signal parameters constant as shown in Fig. 7b. $f_{max}$ values close to 250 GHz are possible at 100 nm channel length. Higher frequency performance will require significant improvements in device parasitics, most notably the contact resistance.

## Acknowledgments

The authors would like to thank K. Watanabe and T. Taniguchi for supplying h-BN crystals. The authors acknowledge the support of the C2S2 Focus Center, one of six research centers funded under the Focus Center Research Program (FCRP), a Semiconductor Research Corporation entity, and DARPA under contract FA8650-08-C-7838 through the CERA program, and by the AFOSR MURI Program on new graphene materials technology, FA9550-09-1-0705.